\documentclass[%
 aip,
 jmp,%
 amsmath,amssymb,
 reprint,%
]{revtex4-1}

\usepackage{graphicx}
\usepackage{dcolumn}
\usepackage{bm}
\usepackage[T1,T2A]{fontenc}
\usepackage[utf8]{inputenc}
\usepackage{CJKutf8}
\usepackage[russian,english]{babel}
\usepackage[overlap, CJK]{ruby}
\usepackage{CJKulem}
\newenvironment{Japanese}{%
  \CJKfamily{min}%
  \CJKtilde
  \CJKnospace}{}

\begin{document}

\begin{CJK}{UTF8}{}

\title{Investigation of the Cubic Boron Nitride Nucleation under the High Pressure and the High Temperature}
\author{A. Su-Gon Kim}
\author{B. Yong-Sob Ri}
\author{C. Gwang-Il Jon}
\author{D. Song-Jin Im}
\affiliation{Department of Physics, Kim Il Sung University, Daesong District, Pyongyang, DPR Korea}

\begin{abstract}
In this paper we have theoretically found the activation energy($420.38kJ/mol$) for the transformation from hBN to cBN in the microscopic viewpoint. We have introduced an analytical formula representing the dependence of nucleus formation time on the activation energy, synthesis pressure and temperature. We have theoretically determined the boundary line of cBN nucleus formation region in the $P-T$ phase diagram on the basis of the diffusion mechanism of cBN nucleus formation. We have found that the cBN crystal nucleus formation time is less than $300ms$ by comparing of the theory and experiment.
\end{abstract}



\maketitle

\section{Introduction}
Cubic boron nitride (cBN) is usually produced in a high pressure-high temperature process by using mostly the hexagonal boron nitride(hBN) as the starting material. 

After the first discover of the cubic boron nitride synthesis method, so far, many studies on the cubic boron nitride synthesis was progressed but the mechanism of the synthesis reaction is still under discussion. Most experimental results  described so far\cite%
{1,2,3,4,5,6,7} suffer from one clear disadvantage. 

Since all experiments for cBN synthesis were carried out in a closed high pressure chamber, they were not observed in situ. Only the final products after quench and pressure release were analyzed and only these results can be taken into account to model the transition mechanism. But the final process of pressure and temperature release may have an important influence. If the products can only be analyzed after experiment, metastable phases may occur or vanish or other phases could undergo phase transition thus leading to incorrect conclusions about the appropriate mechanism. These disadvantages are avoided if the observations are carried out under in situ conditions. In the paper \cite%
{8} by using the synchrotron radiation for energy dispersive X-ray diffraction the cBN synthesis process was observed in situ. It is found that cBN grows directly from an amorphous matrix and not from different BN polymorphs as previously proposed. This is the evidence that the cBN formation process is carried out by the diffusion mechanism.

In this paper, on the basis of these experimental facts, the diffusion mechanism of cBN synthesis were theoretically studied and the cBN synthesis region in the $P-T$ phase diagram was determined.

\section{The Calculation of the Activation Energy}

The individual atom of hBN amorphous phase moves separately to form $sp^3$ bond with cBN crystal particle by diffusion, which would be overcome the energy barrier(activation energy).

$P_{+}(n)$ and $P_{-}(n)$ that are called the probability for jumping of an atom from hBN particle to cBN particle containing the $n$ atoms and leaving of an atom from cBN particle to hBN particle are represented as follows\cite%
{9}

\begin{equation}
P_{\pm}(n)=\tilde{n} \omega exp\left(-\frac{U}{kT}\mp\frac{1}{2kT}\cdot\frac{\partial\Delta U}{\partial n}\right),
\label{eq:1}
\end{equation}
where $\tilde{n}$ and $\omega$ are the number and frequency of atoms in the surface of cBN particle respectively, $\Delta G(n)$ is the change of the Gibbs free energy by forming cBN crystal nucleus containing $n$ atoms. $k$ is Boltzman constant, $U$ is the height of the energy barrier that would be overcame for jumping of an atom from hBN to cBN.

From now, let us estimate the activation energy, height of energy barrier that would be overcome for an atom to break away from $sp^2$ state of hBN to infinitely far dwelling site.

The activation energy $U$ was calculated using expression (\ref{eq:2})

\begin{equation}
U=E_{0}-E_{1}-(E_{B}+E_{N})/2,
\label{eq:2}
\end{equation}
where $E_{0}$ and $E_{1}$ are the energies of hBN perfect crystal and crystal having an vacancy occurred by the diffusion of atom, $E_{B}$ and $E_{N}$ are the spinpolarized energies for the isolated boron and nitrogen atoms. For calculating the energies of the isolated boron and nitrogen atoms, bcc crystal lattice of which B atoms are placed in the apices and a N atom is placed in the middle of a cube with $2.5nm$ of edge length is introduced and local density approximation (LDA) is used. The calculation of energy of crystal having an vacancy is performed by using 16 atoms and relaxation process. The precision of energy is 0.0001eV.

 The activation energy, calculated by using abinit application program according to the above method, is 
\begin{equation}
U=420.38kJ/mol.
\label{eq:3}
\end{equation}

\section{The Kinetic Theory of cBN Crystal Nucleus Formation on the basis of the Diffusion Mechanism}

Now let us consider the kinetic theory of cBN crystal nucleus formation on the basis of the diffusion mechanism. Let us denote the cBN crystal nucleus number consisting of n atoms in $t$ moment by $f (n,t)$. Then the total number of cBN crystal nucleus is as following

\begin{equation}
N(P,T,t)=\sum_{n=0}^\infty f(n,t).
\label{eq:4}
\end{equation}
The dependence of the $f(n,t)$ on the time must satisfy the following equation
\begin{eqnarray}
\frac{\partial f(n,t)}{\partial t}=f(n-1,t)P_{+}(n-1)+\nonumber\\
\: f(n+1,t)P_{-}(n+1)-f(n,t)[P_{+}(n)+P_{-}(n)],
\label{eq:5}
\end{eqnarray}
where $P_{\pm}(n)$ is the probability for jumping of an atom from hBN to cBN and leaving of an atom from cBN to hBN given by formula (\ref{eq:1}) respectively. If $n \gg 1$, equation(5) is approximated as following
\begin{eqnarray}
\frac{\partial f(n,t)}{\partial t}=\frac{\partial}{\partial n}\left\{[P_{-}(n)-P_{+}(n)]f(n,t)\right\}+\nonumber\\
\: \frac{1}{2}\cdot\frac{\partial^2}{\partial n^2}\left\{[P_{+}(n)+P_{-}(n)]f(n,t)\right\}.
\label{eq:6}
\end{eqnarray}
It is the Fokker-Planck equation. However, in generally since $U(n)\gg\frac{\partial \Delta G}{\partial n}$, can be approximated as following

\begin{gather}
P_{-}(n)-P_{+}(n)\cong\tilde{n}\omega e^{-\frac{U}{kT}}\cdot\frac{1}{kT}\frac{\partial\Delta G}{\partial n}
\label{eq:7}\\
P_{+}(n)+P_{-}(n)\cong 2\tilde{n}\omega e^{-\frac{U}{kT}}
\label{eq:8}\\
P(n)\equiv\tilde{n}\omega e^{-\frac{U}{kT}}.
\label{eq:9}
\end{gather}
Using (7), (8)and (9), equation (6) is became as following

\begin{eqnarray}
\frac{\partial f(n,t)}{\partial t}=\frac{\partial^2}{\partial n^2}\left[P(n)f(n,t)\right]+\nonumber\\
\: \frac{1}{kT}\frac{\partial}{\partial n}\left[P(n)f(n,t)\frac{\partial\Delta G}{\partial n}\right].
\label{eq:10}
\end{eqnarray}
In the case $n\gg 1$, since

\begin{equation}
\frac{\partial P(n)}{\partial n}\ll P(n)
\label{eq:11}
\end{equation}
consequently, equation(10) is became as follows

\begin{eqnarray}
\frac{\partial f(n,t)}{\partial t}=\frac{\partial}{\partial n}\left[P(n)\frac{\partial f(n,t)}{\partial n}\right]+\nonumber\\
\: \frac{1}{kT}\frac{\partial}{\partial n}\left[P(n)f(n,t)\frac{\partial \Delta G}{\partial n}\right].
\label{eq:12}
\end{eqnarray}

For quantitatively discussing on the kinetic theory of the cBN crystal nucleus formation, we must solve equation(12) under a given initial and boundary conditions. Since the crystal nucleus does not exist in initial time, we can have the initial condition as following $f(n,0)=0$. In the moment of the crystal nucleus formation with radius of $r$, the Gibbs free energy change is as following

\begin{equation}
\Delta G=\frac{4\pi}{3}r^3n_{0}(\mu_{c}-\mu_{h})+4\pi r^2(\sigma_{c}-\sigma_{h}),
\label{eq:13}
\end{equation}
where $\mu_{c}$ and $\mu_{h}$ are the chemical potential of the cBN and hBN while $\sigma_{c}$  and $\sigma_{h}$ are the surface free energy density of cBN and hBN. $n_{0}$  is the atomic number density of the cBN crystal. Since the crystal nucleus with small size must generate with big probability at a condition $r\to 0$ by the thermodynamic fluctuation theory, we can take the boundary condition as follows

\begin{equation}
\lim_{r \rightarrow 0}\frac{f(r,t)}{f_{e}(r)}\rightarrow 0,
\label{eq:14}
\end{equation}
where $f_{e}(r)$ is the distribution function of the equilibrium state and is as follows

\begin{equation}
f_{e}(r)=Nexp\left(-\frac{\Delta G}{kT}\right),
\label{eq:15}
\end{equation}
where $\Delta G$ is the Gibbs free energy change given by formula(13).

Then the distribution function $f_{e}(r)$ increase infinitely in out a critical size but the distribution function $f(r,t)$ finite so that at case of the condition $r \rightarrow \infty$ we can take the boundary condition as following

\begin{equation}
\lim_{r \rightarrow \infty}\frac{f(r,t)}{f_{e}(r)}\rightarrow 0.
\label{eq:16}
\end{equation}
Let us solve equation (12) on cBN crystal nucleus of the critical size neighborhood under these initial and boundary conditions. In the case that we consider a problem in the critical size neighborhood of cBN crystal nucleus, we can expand $\Delta G(n)$ in the neighborhood of critical size $n_{k}$ by Taylor series and restrict to a second term. Since most of the non-steady state are near to the stationary state, generally, for the rough estimate of the non-steady time we can exchange approximately as following

\begin{eqnarray}
\frac{\partial}{\partial n}\left(P\frac{\partial f}{\partial n}\right) \rightarrow \frac{\partial}{\partial n}\left(P\frac{\partial f_{\infty}}{\partial n}\right)\\
\frac{\partial}{\partial n}\left(Pf\right) \rightarrow \frac{\partial}{\partial n}\left(Pf_{\infty}\right),
\end{eqnarray}
where $f_{\infty}$ is the distribution function of the stationary state that reach through the non-steady process. This stationary state distribution function is given by solving the stationary state Fokker-Planck equation as following

\begin{equation}
\frac{d}{dn}\left(P(n)\frac{df_{\infty}}{dn}\right)+\frac{1}{kT}\frac{d}{dn}\left(P(n)f_{\infty}\frac{d\Delta G}{dn}\right)=0.
\label{eq:19}
\end{equation}
If we transform the equation (12) in consideration of the above facts then the equation (12) is transformed to following simple form

\begin{equation}
\frac{\partial f}{\partial t} \cong \frac{P}{kT}\left(\frac{\partial^2\Delta G}{\partial n^2}\right)_{n=n_{k}}(f-f_{\infty}).
\label{eq:20}
\end{equation}

Solving the equation (20) under initial and boundary conditions, we can obtain the following 
distribution function of cBN crystal nucleus with critical radius

\begin{equation}
f=f_{\infty}\left(1-e^{-t/\tau}\right),
\end{equation}
where $f_{\infty}$ is the stationary state distribution function and $\tau$ is as following

\begin{equation}
\tau=\frac{18ln10\sigma kTe^\frac{U}{kT}}{\pi a_{0}^4\omega n_{0}^2(\Delta \bar{V})^2(P-a-bT)^2},
\end{equation}
where $a_{0}$ is radius of the volume space occupied by an atom, $\Delta\bar{V}=V_{h}-V_{c}$ is the atomic volume difference between hBN and cBN, $a=-1.46GPa, b=0.0031GPa/K$ are constants determining the phase equilibrium line between cBN and hBN and $\omega$  is the oscillation frequency of an atom in the cBN crystal surface, $k$ is Boltzmann constant, $P$ is the pressure,$T$ is the temperature, $U$ is activation energy, $\sigma$ is the surface energy density between cBN crystal nucleus and hBN phase and it can place $\sigma=\sigma_{c}-\sigma_{h}$. As shown from formula (21), since the cBN crystal nucleus distribution function is arrived to the stationary state after $\tau$ time,$\tau$ is called non-steady time and we can call this time as crystal nucleus formation time.

\section{The Determination of the cBN Crystal Nucleus Formation Region}

Representing the formula (22) in terms of the pressure, the following formula is obtained.

\begin{equation}
P=a+bT+\frac{3(2ln10)^{1/2} (\sigma kT)^{1/2} e^{\frac{U}{2kT}}}{(\pi \omega \tau)^{1/2} a_{0}^2 \vert n_{0}\Delta \bar{v}\vert}.
\label{eq:23}
\end{equation}

\begin{figure}[ht]
\includegraphics[width=8cm]{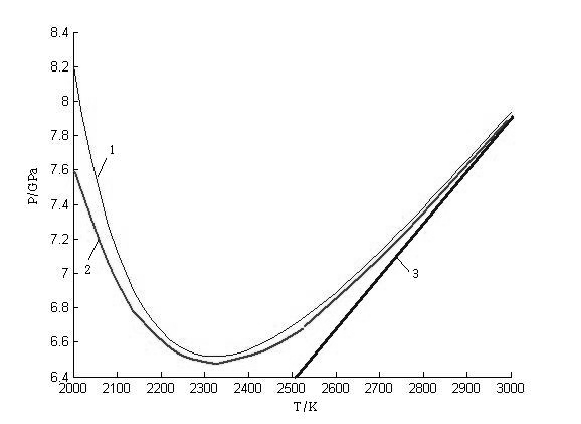}
\caption{cBN nucleation region by the direct transition experiment \cite%
{11} 1 - theory, 2 - experiment, 3 - hBN-cBN phase equilibrium line}
 \end{figure} 
 
Using the formula (\ref{eq:23}), we can find the cBN formation region in the $P-T$ diagram.As known from formula (23), the $P-T$ region in which cBN crystal nucleus is formed is determined by the activation energy and nucleus formation time. For the activation energy we have used expression (2). And the other material constants were equal as following \cite%
{10}. $a_{0}=1.12\cdot 10^{-10} m, n_{0}=1.7\cdot 10^{29} /m^3, V_{c}=5.9\cdot 10^{-30} m^3, V_{h}=9.05\cdot 10^{-30} m^3, \sigma_{c}=4.72J/m^2, \sigma_{h}=3.40J/m^2, \omega \approx 2.0\cdot 10^{14}/s$.

Now we can obtain the boundary line of cBN nucleus formation region in $P-T$ diagram by using formula (22) according to change of the crystal nucleus formation time $\tau$ . However, if $\tau$ is 300ms, theoretical line(curve 1 of Figure 1) coincides with experimental line (curve 2 of Figure 1). Therefore we can estimate cBN crystal nucleus formation
time as less than 300ms.

\section{Conclusion}

In this paper the activation energy for the transformation from hBN to cBN was theoretically found from microscopic viewpoint and estimated to $420.38kJ=mol$. We have introduced an analytical formula representing the dependence of cBN nucleus formation time on the activation energy, synthesis pressure and temperature. Using this formula, the boundary line of cBN nucleus formation region in the $P-T$ phase diagram theoretically was determined. We have found that the crystal
nucleus formation time of cBN is less than $300ms$ by comparing of our theoretical calculation and experiment.
In this paper the temperature effect was not considered. If the temperature effect is considered, then the activation energy will be decreased, so that the crystal nucleus formation time of cBN will be decreased.

\end{CJK}
\end{document}